\documentstyle [12pt,eqsecnum,aps,amsfonts] {revtex}
\input epsf
\newcommand{\beq}{\begin{equation}}
\newcommand{\eeq}{\end{equation}}
\newcommand{\beqs}{\begin{eqnarray}}
\newcommand{\eeqs}{\end{eqnarray}}

\begin{document}

\baselineskip 6.0mm

\begin{center}

{\large\bf End Graph Effects on Chromatic Polynomials for Strip Graphs of 
Lattices and their Asymptotic Limits}

\vspace{4mm}

        Shan-Ho Tsai$^1$

\vspace{6mm}

Institute for Theoretical Physics  \\ 
State University of New York       \\
Stony Brook, N. Y. 11794-3840

\vspace{4mm}

{\large\bf Abstract} 

\end{center} 
\small{
We report exact calculations of the ground state degeneracy per site 
(exponent of the ground state entropy) $W(\{G\},q)$ of the $q$-state Potts 
antiferromagnet on infinitely long strips with specified end graphs for free
boundary conditions in the longitudinal direction and free and periodic
boundary conditions in the transverse direction.  This is equivalent to 
calculating the chromatic polynomials and their asymptotic limits for these 
graphs.  Making the generalization from $q \in {\mathbb Z}_+$ to 
$q \in {\mathbb C}$, we determine the full locus ${\cal B}$ on which $W$ is 
nonanalytic in the complex $q$ plane.  We report the first example for this
class of strip graphs in which ${\cal B}$ encloses regions even for planar 
end graphs.  The bulk of the specific strip graph that exhibits this property
is a part of the $(3^3 \cdot 4^2)$ Archimedean lattice. 

\vspace{10mm}

\begin{flushleft}

1 \  email: tsai@insti.physics.sunysb.edu; address as of May, 1998: Center for 
Simulational Physics, Department of Physics and Astronomy, University of 
Georgia, Athens, GA 30602-2451 \\

\end{flushleft}

\pagebreak

\newpage

\pagestyle{plain}
\pagenumbering{arabic}
\renewcommand{\thefootnote}{\arabic{footnote}}
\setcounter{footnote}{0}

\section{Introduction}

\normalsize

This paper is a continuation of our earlier Ref. \cite{strip} and a 
companion paper to Ref. \cite{strip2} on studies of the ground state degeneracy
per site, $W(\{G\},q)$ (exponent of the ground state entropy $S_0=k_B\ln W$) 
for the $q$-state Potts antiferromagnet (PAF) \cite{potts,wurev} 
on infinitely long strip graphs with specified end 
subgraphs, free boundary conditions in the longitudinal direction, and free 
and periodic boundary conditions in the transverse direction. 
The Potts antiferromagnet is of interest for these
studies because it is a simple model that exhibits nonzero ground state 
entropy on a given lattice for sufficiently large values of $q$.
The zero-temperature partition function of the $q$-state Potts AF on a graph
$G$ is equivalently written as 
$Z(G,q,T=0)_{PAF}=P(G,q)$, where $P(G,q)$ is the chromatic polynomial
\cite{rtrev}, giving the number
of ways of coloring the $n$-vertex graph $G$ with $q$ colors such that no
adjacent vertices have the same color.  Hence, the ground state degeneracy per
site is $W(\{G\},q) = \lim_{n \to \infty}P(G,q)^{1/n}$
where $\{G\}$ denotes the $n \to \infty$ limit of the family of graphs $G$
(e.g., a regular lattice, $\{G\} = \Lambda$ with some specified boundary
conditions).  As before, we shall consider the generalization of $q$ from a
postive integer to a complex quantity.  In general, 
$W(\{G\},q)$ is an analytic function in the $q$ plane
except along a certain continuous locus of points, which we denote ${\cal B}$.

In Ref. \cite{strip} we presented exact
calculations of chromatic polynomials and the asymptotic limiting $W$ functions
for strip graphs of a variety of regular lattices (with free boundary
conditions). For this purpose we developed and applied a generating function 
method. Specifically, we considered strip graphs of regular lattices of type 
$s$ of the form 
\beq
(G_s)_{m;I} = (\prod_{\ell=1}^m H)I
\label{stripi}
\eeq
where $s$ stands for square, triangular, honeycomb, etc.  Thus, 
such a graph is composed of $m$ repetitions of a subgraph $H$ 
attached to an initial subgraph $I$ on one end, which, by convention, we take 
to be the right end.  In the case of homogeneous strips, $I=H$. As is implicit
in 
the above notation, we picture such strip graphs as having a length of $L_x$ 
vertices in the horizontal ($x$) direction and a width of $L_y$ vertices in the
transverse (vertical, $y$) direction.  Some further related work is in Ref. 
\cite{hs}. 

In Ref. \cite{strip2} we generalized the study in Ref. \cite{strip} by 
considering strip 
graphs of regular lattices in which there may be special subgraphs on both 
the left and right ends of the strip: 
\beq
(G_s)_{m,J,I} = J(\prod_{\ell=1}^m H)I
\label{stripij}
\eeq
The end subgraphs $I$ and $J$ may be different from each other and from 
the repeated subgraph $H$.  Furthermore, while the boundary conditions in the
longitudinal (horizontal) direction were still free, we studied both free 
and periodic boundary conditions in the transverse (vertical) direction.  
We found that while strips of the form (\ref{stripi}) yield loci ${\cal B}$
that consist of arcs and possible line segments which do not depend on the end
graph $I$ and which do not enclose regions, strips of the form (\ref{stripij})
yield loci ${\cal B}$ that do, in general, depend on the end graphs $I$ and $J$
and in certain cases do enclose regions.  However, for the strips and end
graphs considered in Ref. \cite{strip2}, ${\cal B}$ only enclosed regions for
end graphs that were nonplanar and hence were fundamentally different from the
bulk (interior) of the strips, which was planar for both free and periodic
boundary conditions in the transverse direction.  
In the present paper we shall report on further work on this subject and, in
particular, our finding of the first case, for this class of strip graphs, 
in which ${\cal B}$ encloses regions for end graphs that are planar, like 
the rest of the strip.  As we shall show, this occurs for strips of a specific
type of Archimedean lattice, commonly denoted in the mathematical literature by
the symbol $(3^3 \cdot 4^2)$.  Here we recall the definition of an Archimedean
lattice as a uniform tiling of the plane by one or more regular polygons, such
that each vertex is equivalent to every other vertex \cite{gs}.  Thus an
Archimedean lattice $\Lambda$ is 
defined by the ordered sequence of polygons that one traverses in
making a complete circuit around a vertex in a given (say counterclockwise)
direction: 
\beq
\Lambda = (\prod_i p_i^{a_i})
\label{archlambda}
\eeq
where in the above circuit, the notation $p_i^{a_i}$ indicates that the
regular $p$-sided polygon $p_i$ occurs contiguously $a_i$ times (it can also
occur noncontiguously).  In this paper we shall concentrate on a particular
heteropolygonal Archimedean lattice, symbolized in the notation of
eq. (\ref{archlambda}) as 
\beq
\Lambda_{33344} = (3^3 \cdot 4^2)
\label{lam33344}
\eeq
An illustration of strips of this lattice is shown in Fig. \ref{strip34}.  
Evidently, this lattice has
coordination number $\Delta =5$, girth $g=3$ (where the girth is the 
length of minimum-length nontrivial closed circuit on the lattice) and
chromatic number $\chi=3$. (The chromatic number $\chi(G)$ of a graph 
$G$ is defined as the minimum number of colors, $q \in {\mathbb Z}_+$ which 
are necessary to color the graph such that no two adjacent vertices have the 
same color, i.e., the minimum $q \in {\mathbb Z}_+$ such that $P(G,q) > 0$.)
We have obtained rigorous bounds on the ground state entropy of the Potts
antiferromagnet on this lattice \cite{wn}. 
\begin{figure}
\centering
\leavevmode
\epsfxsize=3.5in
\epsffile{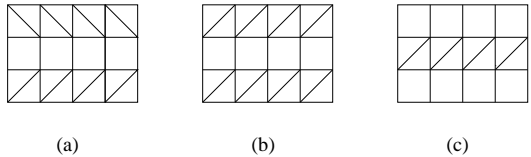}
\caption{\footnotesize{Illustrations of strips of the $(3^3 \cdot 4^2)$ 
lattice of width $L_y=4$. We consider the three types of strips shown in 
(a), (b) 
and (c), which we shall refer to as strip graphs of type $(33344a)$, 
$(33344b)$ and $(33344c)$, respectively. These illustrations show strips with 
$L_x=m+2=5$, corresponding to $n=20$ vertices.}}
\label{strip34}
\end{figure}

For some relevant definitions and notation, the reader is referred to Refs. 
\cite{wn} and \cite{p3afhc}-\cite{ww}.  In particular, we define the
maximal finite real point where $W$ is nonanalytic as $q_c$.  
It will also be convenient to define the function 
\beq
D_k(q) = \sum_{s=0}^{k-2}(-1)^s {{k-1}\choose {s}} q^{k-2-s}
\label{dk}
\eeq
Some further details are given in Ref. \cite{thesis}.  
This paper is organized as follows. In Section II we introduce
our notation and briefly summarize the method of generating functions used
here. In Section III we present generating functions and the
analytic structure of the associated $W$ functions for strips of the 
$(3^3\cdot 4^2)$ lattice with free boundary conditions in the longitudinal and
transverse directions of the strips. Section IV reports some
results for strips of the $(3^3\cdot 4^2)$ lattice with free boundary 
conditions in the longitudinal direction and periodic boundary conditions in
the transverse direction of the strips. Section V contains some conclusions. 

\vspace{14mm}

\section{Generating Function Method and End Subgraphs }

   We denote the chromatic polynomial describing the coloring of the strip
graph $(G_s)_{m,J,I}$ with $q$ colors as $P((G_s)_{m,J,I},q)$.  In the
generating function method, this chromatic polynomial is given by means of an
expansion, in terms of an auxiliary variable $x$, of a generating function
$\Gamma(G_{s,J,I},q,x)$: 
\beq
\Gamma(G_{s,J,I},q,x) = \sum_{m=0}^{\infty} P((G_s)_{m,J,I},q)x^m
\label{gamma}
\eeq
where $\Gamma(G_{s,J,I},q,x)$ is a rational function of the form
\beq
\Gamma(G_{s,J,I},q,x) = \frac{{\cal N}(G_{s,J,I},q,x)}
{{\cal D}(G_{s,J,I},q,x)}
\label{gammagen}
\eeq
with (suppressing $J,I$ dependence in the notation)
\beq
{\cal N}(G_s,q,x) = \sum_{j=0}^{j_{max}} a_{G_s,j}(q) x^j
\label{n}
\eeq
and
\beq
{\cal D}(G_s,q,x) = 1 + \sum_{k=1}^{k_{max}} b_{G_s,k}(q) x^k
\label{d}
\eeq
Our calculational methods are described in Refs. \cite{strip,strip2}. 
Equivalently to eq. (\ref{gamma}), we calculate the chromatic polynomial as 
\beq
P((G_{s,J,I})_m,q)={\bf v_b}^T (MD)^m M {\bf v_a}
\label{addconpol}
\eeq
where $D$ and $M$ are $L \times L$ matrices, with $L$ depending on the width of
the strip.  Here, as in Ref. \cite{strip2}, we shall restrict ourselves to the
cases where $L=5$ for strips with free boundary conditions in the longitudinal
and tranverse directions and to the cases where $L=4$ for strips with
free boundary condition in the longitudinal direction and periodic boundary
condition in the transverse direction, since these are sufficient to exhibit 
the new phenomenon that we wish to discuss. For the case $L=5$, the matrix $D$
is given by 
\beq
D =  \left (\begin{array}{ccccc}
\frac{1}{q(q-1)(q-2)(q-3)} & 0 & 0 & 0 & 0 \\
0 & \frac{1}{q(q-1)(q-2)} & 0 & 0 & 0 \\
0 & 0 & \frac{1}{q(q-1)(q-2)} & 0 & 0 \\
0 & 0 & 0 & \frac{1}{q(q-1)} & 0 \\
0 & 0 & 0 & 0 & \frac{1}{q(q-1)(q-2)} \end{array}\right )
\label{dmatrix}
\eeq
whereas for the case $L=4$ the matrix $D$ is the upper left-hand $4 \times 4$ 
submatrix of $D$ given above in eq. (\ref{dmatrix}). 
The matrix $M$ depends on the specific strip.  
Factorizing the denominator of the generating function (\ref{d}), we have 
\beq
{\cal D}(G_{s,J,I},q,x)=\prod_r (1-\lambda_r(q) x),
\label{addconden}
\eeq
where the 
${\lambda_r(q)}'s$ are eigenvalues of the product of matrices $MD$. The
subgraphs on the two ends of the strip (defined by ${\bf v_a}$ and
${\bf v_b}$) determine which eigenvalues enter in the product in equation
(\ref{addconden}).  Using this method, we have calculated chromatic
polynomials and their asymptotic limits for various strip graphs of the form
$(G_s')_m \equiv (G_{s,J,I})_m = J(\prod_{\ell=1}^m H)I$, with $(G_s')_0 
\equiv JI$, where the end graphs $I$ and $J$ are, in general, different. 
We follow the same notation as in Ref. \cite{strip2} for the vectors that 
describe the subgraphs on the ends of the strips: (i) for the case $L=5$,
${\bf v}_a$ and ${\bf v}_b$ will be chosen from the following vectors: 
\beq
{\bf v_1}=(1,0,0,0,0)^T
\label{v1}
\eeq
\beq
{\bf v_2}=(0,1,0,0,0)^T
\label{v2}
\eeq
\beq
{\bf v_3}=(0,0,1,0,0)^T
\label{v3}
\eeq
\beq
{\bf v_4}=(0,0,0,1,0)^T
\label{v4}
\eeq
\beq
{\bf v_5}=(0,0,0,0,1)^T
\label{v5}
\eeq
\beq
{\bf v_6}=(1,1,0,0,0)^T
\label{v6}
\eeq
\beq
{\bf v_7}=(1,0,1,0,0)^T
\label{v7}
\eeq
\beq
{\bf v}=(1,1,1,1,1)^T
\label{v8}
\eeq 
(ii) for the case $L=4$, 
${\bf v}_a$ and ${\bf v}_b$ will be chosen from the following vectors: 
\beq
{\bf v_1}=(1,0,0,0)^T
\label{v1L4}
\eeq
\beq
{\bf v_2}=(0,1,0,0)^T
\label{v2L4}
\eeq
\beq
{\bf v_3}=(0,0,1,0)^T
\label{v3L4}
\eeq
\beq
{\bf v_4}=(0,0,0,1)^T
\label{v4L4}
\eeq
\beq
{\bf v_5}=(1,1,0,0,)^T
\label{v6L4}
\eeq
\beq
{\bf v_6}=(1,0,1,0)^T
\label{v7L4}
\eeq
\beq
{\bf v}=(1,1,1,1)^T
\label{v8L4}
\eeq
For ${\bf v}$, the end subgraph is the same as the transverse set of edges and
vertices. The end graphs corresponding to the vectors in eqs. 
(\ref{v1})-(\ref{v8}) were illustrated in Ref. \cite{strip2}. 

For strip graphs with free boundary conditions in the $x$ and $y$ directions
(FBC$_x$ and FBC$_y$), 
we shall restrict ourselves to cases where the product 
of matrices $MD$ is a $5\times 5$ matrix. 
For strip graphs with free boundary conditions in the $x$ direction (FBC$_x$)
and periodic boundary conditions in the $y$ direction (PBC$_y$) 
we shall restrict ourselves to cases where the product 
of matrices $MD$ are $4\times 4$ matrices. With these preliminaries covered, 
we proceed to our new results. 

\section{Strips of the $(3^3 \cdot 4^2)$ Lattice of Width 
$L_{\lowercase{y}}=4$ and FBC$_{\lowercase{y}}$ }

We consider three types of strips of the $(3^3 \cdot 4^2)$ lattice of width 
$L_y=4$. Sections of these strips are illustrated in Figs. \ref{strip34}(a),
\ref{strip34}(b) and \ref{strip34}(c) and will be referred to as 
strip graphs of type $(33344a)$, $(33344b)$ and $(33344c)$, respectively.

\subsection{Type $(33344{\lowercase{a}})$}

For a strip graph of type $(33344a)$, width $L_y=4$ and end subgraphs  
described by vectors ${\bf v}$, i.e., end subgraphs as shown in Fig. 
\ref{strip34}(a), we 
obtain a generating function of the form of Eqs. (\ref{gammagen}), (\ref{n}) 
and (\ref{d}), with $j_{max}=2$, $k_{max}=3$ and coefficients
\beq
a_{33344a(4),0}= q(q-1)(q^2-3q+3)(q-2)^4,
\label{a0s34aw3}
\eeq
\beq
a_{33344a(4),1}=-q(q-1)(2q^5-21q^4+86q^3-171q^2+164q-59)(q-2)^3,
\label{a1s34aw3}
\eeq
\beq
a_{33344a(4),2}=q(q-3)(q^2-5q+7)(q-1)^3(q-2)^5,
\label{a2s34aw3}
\eeq
\beq
b_{33344a(4),1}=-q^4+9q^3-34q^2+66q-55,
\label{b1s34aw3}
\eeq
\beq
b_{33344a(4),2}=(q-2)(q-3)(2q^2-11q+16)(q^2-4q+5),
\label{b2s34aw3}
\eeq
\beq
b_{33344a(4),3}=-(q-3)(q^2-5q+7)(q-2)^5.
\label{b3s34aw3}
\eeq

Three of the five eigenvalues of $MD$ are the inverses of the three roots
of $1+b_{33344a(4),1} x+b_{33344a(4),2} x^2 +b_{33344a(4),3} x^3$. These 
eigenvalues contribute to the generating functions for all end subgraphs
considered here. The other two eigenvalues are $\lambda_4=1$ and $\lambda_5 
=(q-2)^2$, which only contribute to the generating functions for some
end subgraphs.  In Table \ref{table34a} we show for some illustrative 
planar and non-planar end subgraphs, whether or not $\lambda_4$ and  
$\lambda_5$ contribute to the generating function and some features of the 
locus ${\cal B}$ for each case. 

\pagebreak

\begin{table}
\caption{\footnotesize{
Illustrative end subgraphs for a strip of the $(3^3 \cdot 4^2)$ lattice 
of type $(33344a)$  and width $L_y=4$. 
The notation ${\bf v_J},{\bf v_I}$ means that the vectors ${\bf v}_J$ and 
${\bf v}_I$ describe the subgraphs on the left- and right-hand end of the
strip. P stands for planar and NP for non-planar. The third and fourth 
columns show respectively whether or
not $\lambda_4$ and $\lambda_5$ enter in the generating function. The fifth
column lists some features of ${\cal B}$.}}
\begin{center}
\footnotesize
\begin{tabular}{|ccccc|} \hline
$J,I$ & planarity & does $\lambda_4$ enter? & does $\lambda_5$ enter? 
& features of ${\cal B}$\\
\hline\hline
${\bf v},{\bf v}$     & P,P & N & N & arcs \\
${\bf v},{\bf v_i}$, $i=1,2,3,4$ & P,NP   & N & N & arcs \\
${\bf v},{\bf v_i}$, $i=5,6,7$   & P,P & N & N & arcs \\
${\bf v_1},{\bf v_i}$, $i=1,2,3,4$  & NP,NP & Y & N & arcs and one enclosed 
region\\
${\bf v_1},{\bf v_i}$, $i=5,6,7$    & NP,P & N & N & arcs \\
${\bf v_2},{\bf v_i}$, $i=2,3$ & NP,NP & Y & Y & arcs and three enclosed 
regions\\
${\bf v_2},{\bf v_i}$, $i=6,7$ & NP,P & N & Y & arcs and two enclosed regions\\
${\bf v_3},{\bf v_3}$  & NP,NP & Y & Y & arcs and three enclosed regions \\
${\bf v_4},{\bf v_i}$, $i=2,3,4$ & NP,NP & Y & N & arcs and one enclosed 
region\\
${\bf v_4},{\bf v_i}$, $i=6,7$ & NP,P & N & N & arcs \\
${\bf v_5},{\bf v_i}$, $i=2,3,4$ & P,NP & N & N & arcs \\
${\bf v_5},{\bf v_i}$, $i=5,6,7$ & P,P & N & N & arcs \\
${\bf v_6},{\bf v_i}$, $i=2,3$ & P,NP & N & Y & arcs and two enclosed regions\\
${\bf v_6},{\bf v_i}$, $i=6,7$ & P,P & N & Y & arcs and two enclosed regions\\
${\bf v_7},{\bf v_i}$, $i=2,3$ & P,NP & N & Y & arcs and two enclosed regions\\
${\bf v_7},{\bf v_7}$ & P,P & N & Y & arcs and two enclosed regions\\
\hline
\end{tabular}
\normalsize
\end{center}
\label{table34a}
\end{table}

\pagebreak

\begin{figure}
\centering
\leavevmode
\epsfxsize=3.5in
\epsffile{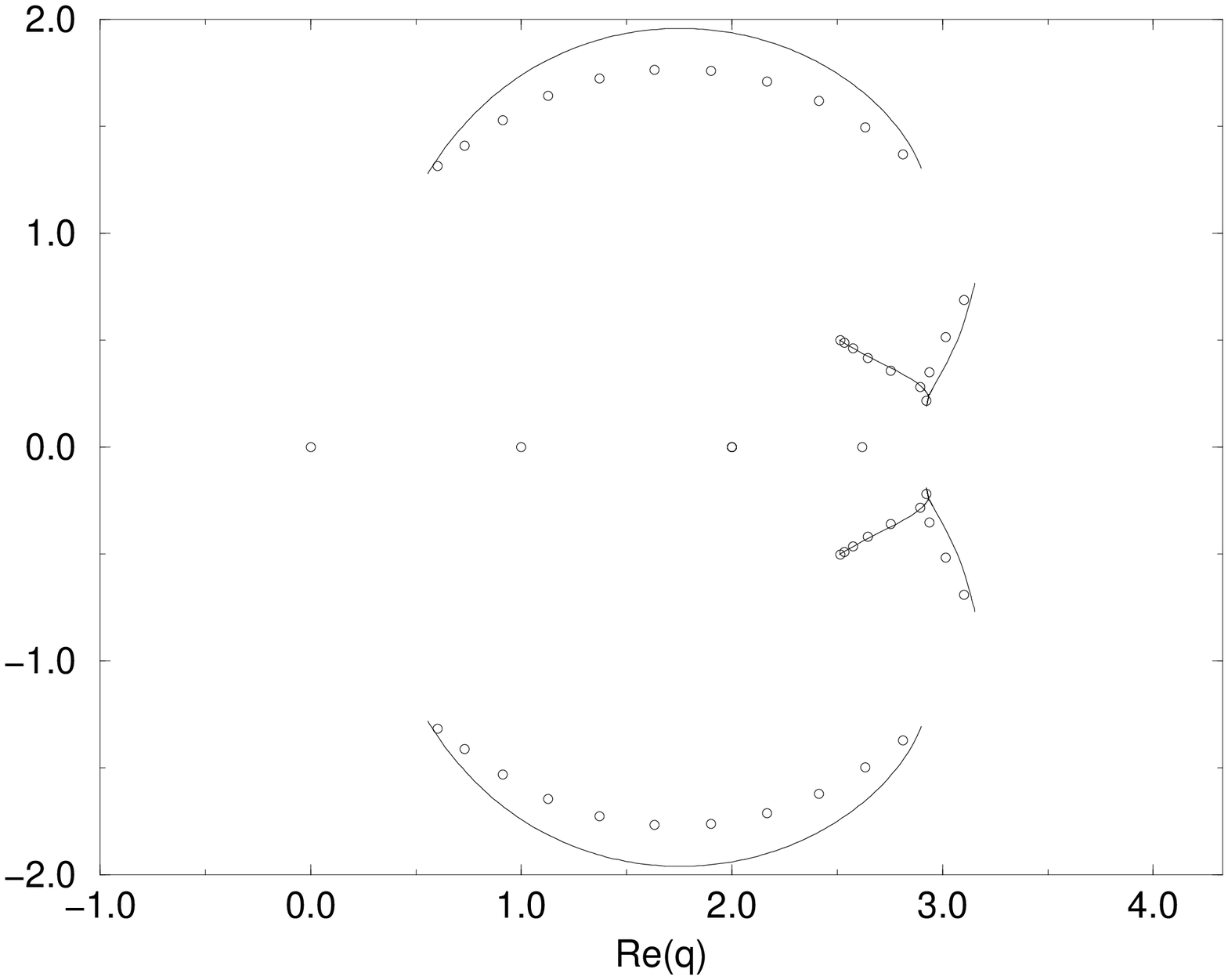}
\caption{\footnotesize{Analytic structure of the function 
$W(\{G_{33344a(4)}'\},q)$ for the strip of the $(3^3 \cdot 4^2)$ lattice of 
type $(33344a)$, width $L_y=4$ and length $L_x=\infty$.
End subgraphs are such that $\lambda_4$ and $\lambda_5$ do not appear 
in the generating function. For comparison, the zeros of the chromatic 
polynomial for a strip with $m=10$ repeating units ($n=48$ vertices) 
and subgraphs 
described by ${\bf v}$ on the right and left ends of the strip, are shown.}}
\label{strip34a0r}
\end{figure}

\pagebreak

\begin{figure}
\centering
\leavevmode
\epsfxsize=3.5in
\epsffile{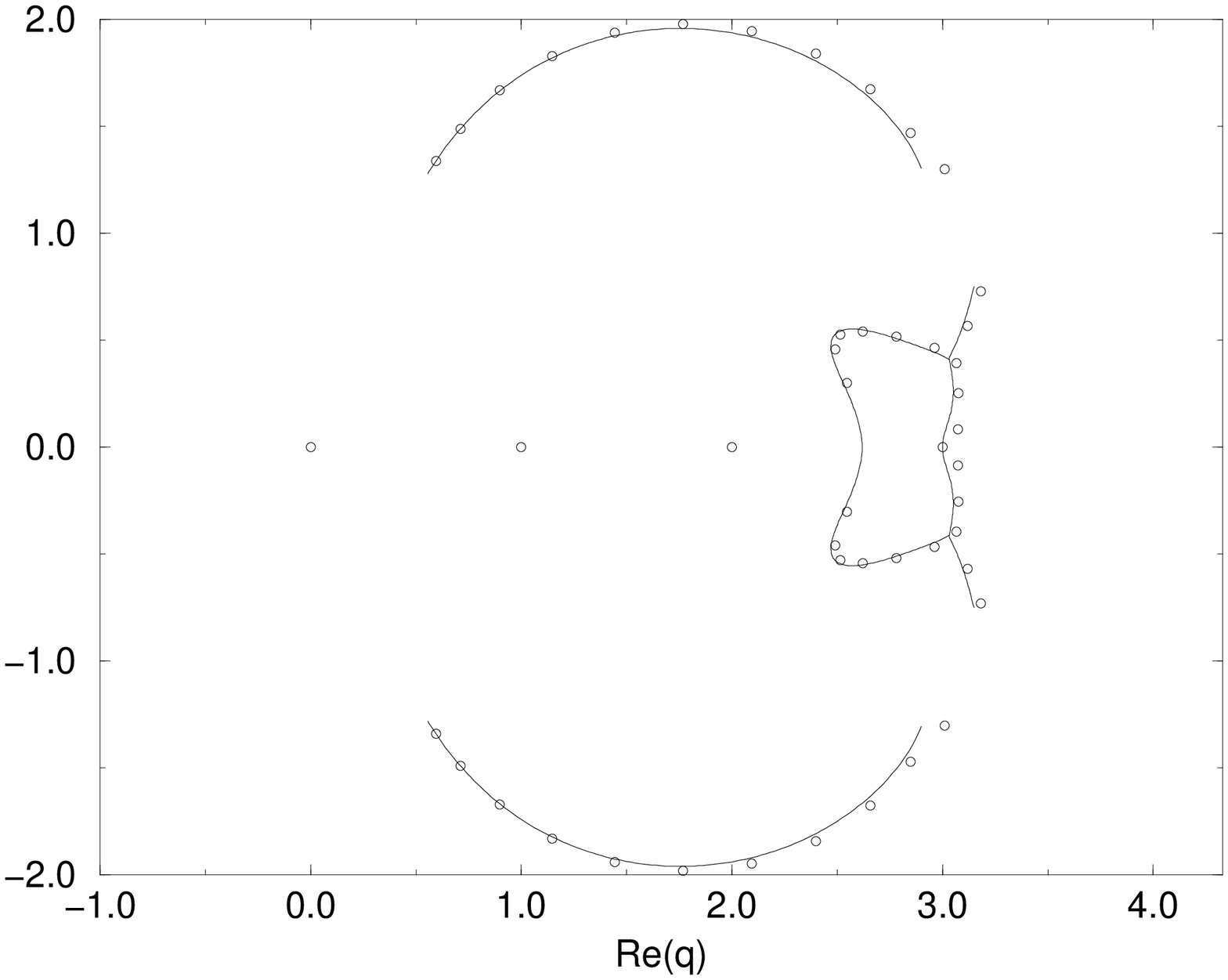}
\caption{\footnotesize{Analytic structure of the function 
$W(\{G_{33344a(4)}'\},q)$ for the strip of the $(3^3 \cdot 4^2)$ lattice of 
type $(33344a)$, width $L_y=4$ and length $L_x=\infty$.
End subgraphs are such that $\lambda_4$ appears 
in the generating function, but $\lambda_5$ does not. For comparison, the 
zeros of the chromatic polynomial for a strip with $m=10$ repeating units 
($n=48$ vertices) and subgraphs 
described by ${\bf v_1}$ on the right and left ends of the strip, are shown.}}
\label{strip34a1r}
\end{figure}

\pagebreak

\begin{figure}
\centering
\leavevmode
\epsfxsize=3.5in
\epsffile{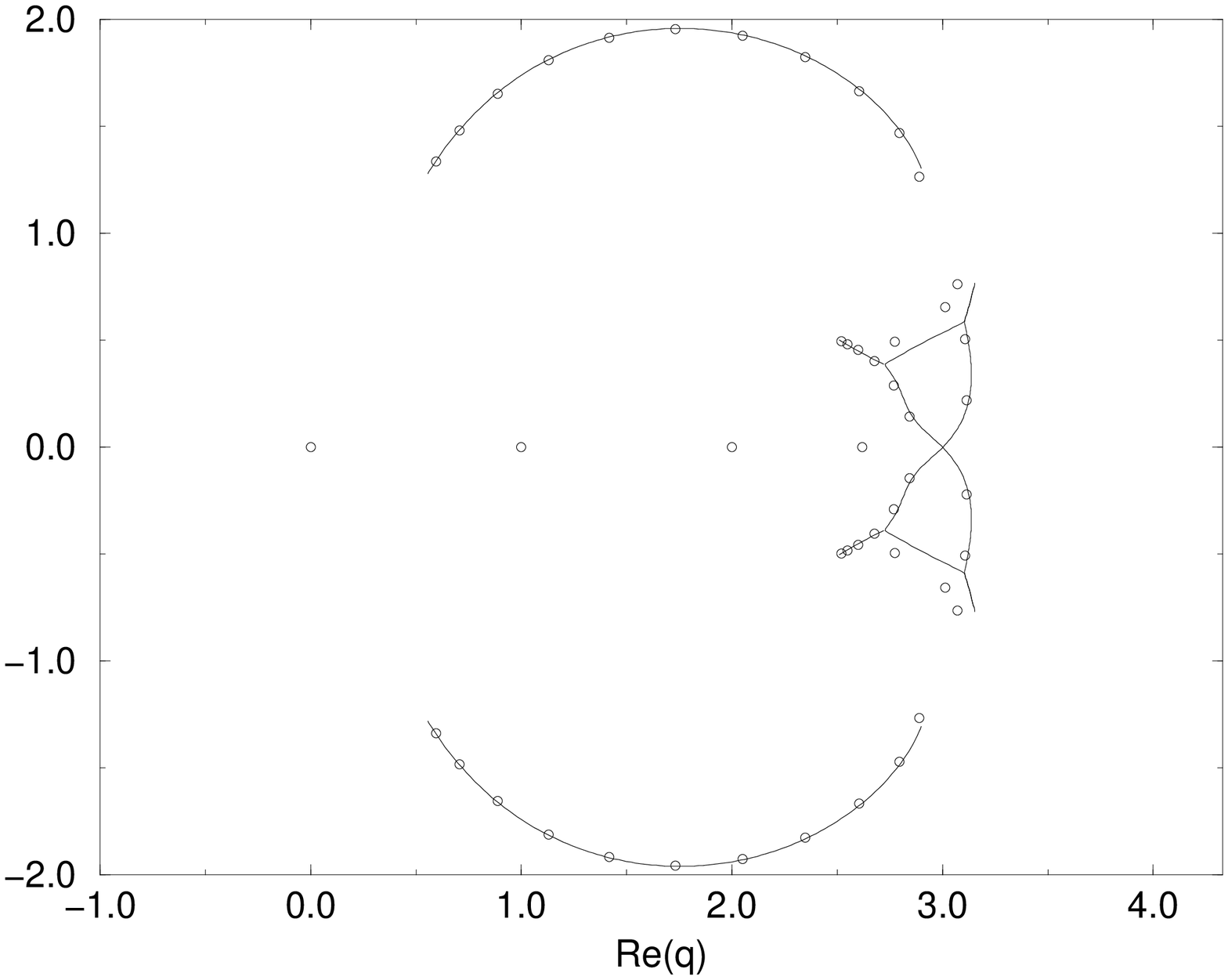}
\caption{\footnotesize{Analytic structure of the function 
$W(\{G_{33344a(4)}'\},q)$ for the strip of the $(3^3 \cdot 4^2)$ lattice of 
type $(33344a)$, width $L_y=4$ and length $L_x=\infty$.
End subgraphs are such that $\lambda_5$ appears 
in the generating function, but $\lambda_4$ does not. For comparison, the 
zeros of the chromatic polynomial for a strip with $m=10$ repeating units 
($n=48$ vertices) and subgraphs 
described by ${\bf v_6}$ on the right and left ends of the strip, are shown.}}
\label{strip34a2r}
\end{figure}

\pagebreak

\begin{figure}
\centering
\leavevmode
\epsfxsize=3.5in
\epsffile{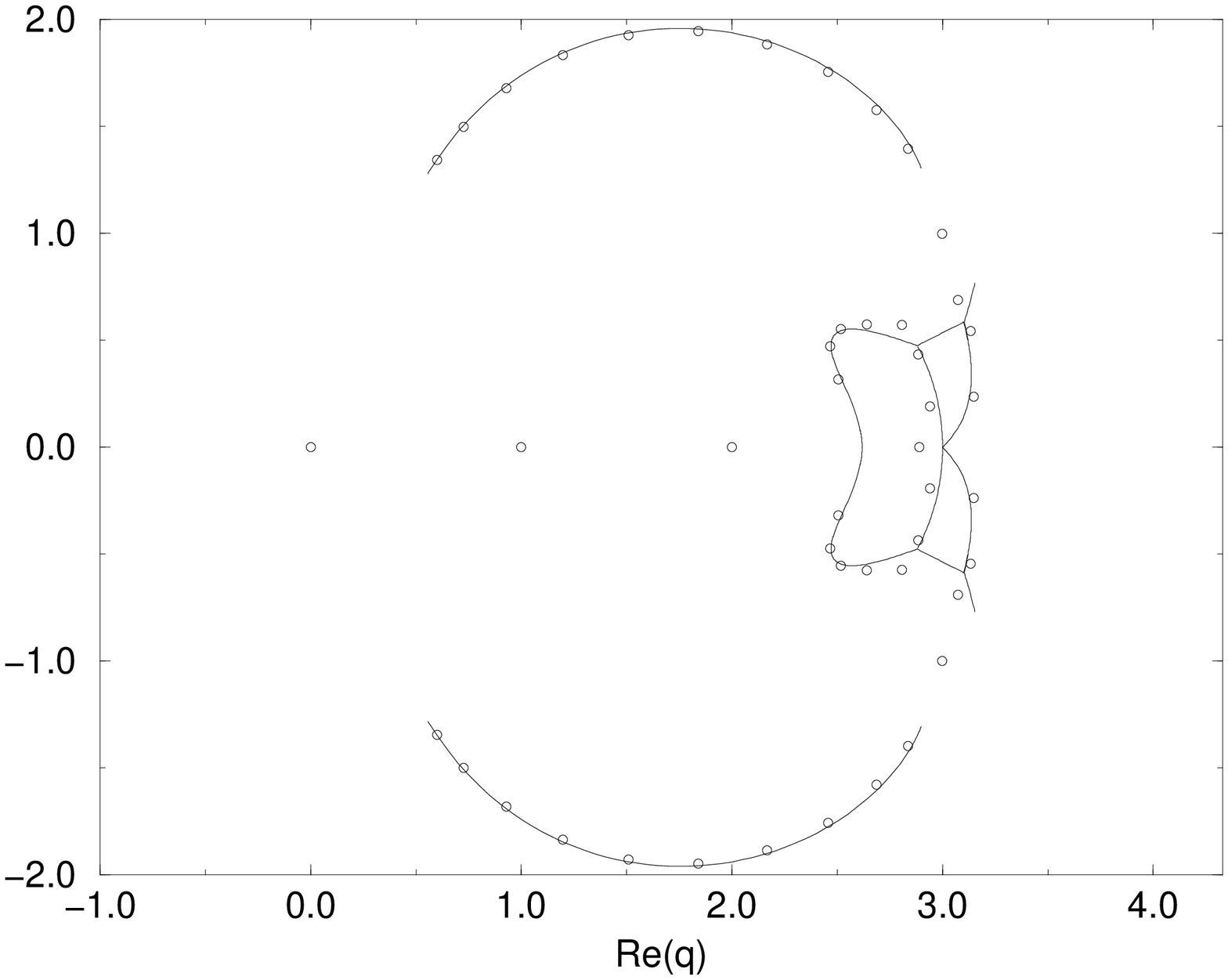}
\caption{\footnotesize{Analytic structure of the function 
$W(\{G_{33344a(4)}'\},q)$ for the strip of the $(3^3 \cdot 4^2)$ lattice of 
type $(33344a)$, width $L_y=4$ and length $L_x=\infty$.
End subgraphs are such that $\lambda_4$ and $\lambda_5$ appear
in the generating function. For comparison, the zeros of the chromatic 
polynomial for a strip with $m=10$ repeating units ($n=46$ vertices) 
and subgraphs described by ${\bf v_2}$ on the right and left ends 
of the strip, are shown. }}
\label{strip34a3r}
\end{figure}

Figs. \ref{strip34a0r}, \ref{strip34a1r}, \ref{strip34a2r} and 
\ref{strip34a3r} show the analytic structure of $W(\{G_{33344a(4)}'\},q)$
for a strip of the $(3^3 \cdot 4^2)$ lattice of type $(33344a)$ and width 
$L_y=4$. Fig. \ref{strip34a0r} corresponds to end subgraphs that 
do not yield $\lambda_4$ and $\lambda_5$ in the generating function. In these
cases the locus ${\cal B}$ is formed by pairs of complex conjugate 
arcs that do not separate the $q$ plane into separate regions.  One can observe
also the features that (i) there are two complex-conjugate multiple points
\cite{alg} on ${\cal B}$ at $q_m \simeq 2.93\pm 0.24i$ and (ii) ${\cal B}$
does not cross the real $q$ axis. These multiple points have the same topology
as the one shown in Fig. \ref{fig34enlarge}(b).  Fig. \ref{strip34a1r} 
corresponds to end subgraphs that yield $\lambda_4$ but do not yield
$\lambda_5$ in the generating function. In these cases the locus ${\cal B}$
is comprised by arcs which enclose one self-conjugate region, where 
$\lambda_4$ is the leading eigenvalue. The locus ${\cal B}$ crosses the real
$q$ axis at $q=\frac{3+\sqrt{5}}{2}=2.618...$ and at $q=3$. Thus $q_c=3$ for 
end subgraphs that yield ${\cal B}$ shown in this figure. 
Fig. \ref{strip34a2r} corresponds to end subgraphs that yield
$\lambda_5$ but do not yield $\lambda_4$ in the generating function. In 
these cases the locus ${\cal B}$ is comprised by arcs which enclose one 
pair of complex-conjugate regions, where $\lambda_5$ is the leading 
eigenvalue. The locus ${\cal B}$ crosses the real $q$ axis at one point,
namely at $q=3$. When the end subgraphs are such that both 
$\lambda_4$ and $\lambda_5$ appear in the generating function, the boundary 
${\cal B}$ is comprised by arcs which enclose one self-conjugate region and 
one pair of complex-conjugate regions, as shown in Fig. \ref{strip34a3r}. In 
the self-conjugate region the leading eigenvalue is $\lambda_4$, whereas in 
the complex-conjugate regions $\lambda_5$ is leading. The locus ${\cal B}$ 
crosses the real $q$ axis at $q=\frac{3+\sqrt{5}}{2}=2.618...$ and at $q=3$.
It is interesting to note that in all cases where ${\cal B}$ crosses the real
$q$ axis, the rightmost crossing point is at $q_c=3$. 

The locus ${\cal B}$  shown in Fig. \ref{strip34a0r} contains 
arc endpoints at $q_{e1}\simeq 0.556+1.281i$, $q_{e2} \simeq 2.898+1.307i$,
$q_{e3} \simeq 3.154 + 0.767i$, $q_{e4} \simeq 2.923+0.190i$, 
$q_{e5} \simeq 2.509+0.502i$ and at the complex-conjugates of these points.
In the locus ${\cal B}$ shown in Fig. \ref{strip34a1r} 
the arc endpoints 
occur at $q_{e1}$, $q_{e2}$, $q_{e3}$ and their complex-conjugates. The points
$q_{e4}$ and $q_{e5}$ (and their complex-conjugates) lie within the region 
where $\lambda_4$ is leading and do not correspond to arc endpoints. The 
locus ${\cal B}$ shown in 
Fig. \ref{strip34a2r} contains arc endpoints at  $q_{e1}$, $q_{e2}$, $q_{e3}$,
$q_{e5}$  and their complex-conjugates. The point $q_{e4}$ and its 
complex-conjugate lie within the regions where $\lambda_5$ is leading and
do not correspond to arc endpoints. The locus ${\cal B}$ shown in 
Fig. \ref{strip34a3r} contains arc endpoints at  $q_{e1}$, $q_{e2}$, $q_{e3}$,
and their complex-conjugates. The points
$q_{e4}$ and $q_{e5}$ (and their complex-conjugates) lie within the region 
where $\lambda_4$ is leading and do not correspond to arc endpoints.

For a strip of the triangular lattice of width $L_y=4$, as discussed in Ref.
\cite{strip2}, we
only obtained loci ${\cal B}$ comprised by arcs separating the complex 
$q$ plane into disconnected regions when subgraphs on the right
and left ends of the strip involved non-planar graphs. For a strip of the 
$(3^3 \cdot 4^2)$ lattice of type $(33344a)$ and width $L_y=4$,
planar subgraphs on one end or on both ends of the strip can yield 
arcs which enclose regions in ${\cal B}$. 

\subsection{Type $(33344b)$}

For a strip graph of type $(33344b)$, width $L_y=4$ and end subgraphs 
described by vectors
${\bf v}$, i.e., as shown in Fig. \ref{strip34}(b), we obtain
the same generating function as for the strip graph $(33344a)$, with 
coefficients given in Eqs. (\ref{a0s34aw3})-(\ref{b3s34aw3}). The five 
eingenvalues of $MD$ are also the same as for the strip graph of type 
$(33344a)$ and width $L_y=4$. 
However, end subgraphs that yield $\lambda_4$ or $\lambda_5$ or
both in the generating functions are not identical for these two types of 
strips. In Table \ref{table34b} we show for some illustrative 
planar and non-planar end subgraphs, whether or not $\lambda_4$ and  
$\lambda_5$ contribute to the generating function and some features of the 
loci ${\cal B}$ for each case. 

\begin{table}
\caption{\footnotesize{
Illustrative end subgraphs for a strip of the $(3^3 \cdot 4^2)$ lattice 
of type $(33344b)$ and width $L_y=4$. 
The notation is the same as in Table \ref{table34a}.}}

\begin{center}
\footnotesize
\begin{tabular}{|ccccc|} \hline
$J,I$ & planarity & does $\lambda_4$ enter? & does $\lambda_5$ enter? 
& features of ${\cal B}$\\
\hline\hline
${\bf v},{\bf v}$     & P,P & N & N & arcs \\
${\bf v_i},{\bf v_i}$, $i=1,4$ & NP,NP   & Y & Y & arcs and three enclosed 
regions \\
${\bf v_i},{\bf v_i}$, $i=2,3$   & NP,NP & Y & N & arcs and one enclosed 
region\\
${\bf v_5},{\bf v_5}$    & P,P & N & Y & arcs and two enclosed regions\\
${\bf v_i},{\bf v_i}$, $i=6,7$    & P,P & N & N & arcs \\
\hline
\end{tabular}
\normalsize
\end{center}
\label{table34b}
\end{table}

For strips of the $(3^3 \cdot 4^2)$ lattice of width $L_y=4$ and types 
$(33344a)$ and $(33344b)$ the analytic struture of $W(\{G_{33344(4)}'\},q)$ 
can have enclosed regions when (i) the subgraphs on one end of the strip 
involve a non-planar graph and on the other end a planar graph, (ii) the
subgraphs on both ends involve planar graphs or (iii) the 
subgraphs on both ends involve non-planar graphs.

\subsection{Type $(33344c)$}

For a strip graph of type $(33344c)$ and subgraphs described by 
vectors ${\bf v}$, i.e., as shown in Fig. \ref{strip34}(c), we 
obtain a generating function of the form of Eqs. (\ref{gammagen}), (\ref{n}) 
and (\ref{d}), with $j_{max}=3$, $k_{max}=4$ and coefficients
\beq
a_{33344c(4),0}=q(q-1)(q-2)^2(q^2-3q+3)^2
\label{a0s34cw3}
\eeq
\beq
a_{33344c(4),1}=-q(q-1)(q-2)(3q^7-38q^6+208q^5-638q^4+1185q^3-1330q^2+829q-217)
\label{a1s34cw3}
\eeq
\beq
a_{33344c(4),2}=q(q-1)(3q-4)(q^2-3q+3)(q^3-7q^2+15q-8)(q-2)^4
\label{a2s34cw3}
\eeq 
\beq
a_{33344c(4),3}=-q(q^2-5q+5)^2(q-1)^3(q-2)^6
\label{a3s34cw3}
\eeq
\beq
b_{33344c(4),1}=-q^4+8q^3-29q^2+56q-46
\label{b1s34cw3}
\eeq
\beq
b_{33344c(4),2}=3q^6-38q^5+208q^4-634q^3+1136q^2-1122q+465
\label{b2s34cw3}
\eeq
\beq
b_{33344c(4),3}=-3q^8+52q^7-395q^6+1720q^5-4702q^4+8264q^3-9101q^2+5718q-1561
\label{b3s34cw3}
\eeq
\beq
b_{33344c(4),4}=(q-2)^6(q^2-5q+5)^2
\label{b4s34cw3}
\eeq

Four of the five eigenvalues of $MD$ are the inverses of the four roots
of $1+b_{33344c(4),1} x+b_{33344c(4),2} x^2 +b_{33344c(4),3} x^3 +
b_{33344c(4),4} x^4$. These 
eigenvalues contribute to the generating functions for all end subgraphs 
considered here. The other eigenvalue is $\lambda_5=1$, which contributes 
to the generating functions for some end subgraphs.
In Table \ref{table34c} we show for some illustrative 
planar and non-planar end subgraphs, whether or not $\lambda_5$
contributes to the generating function and some features of the locus 
${\cal B}$ for each case. 

\begin{table}
\caption{\footnotesize{
Illustrative end subgraphs for a strip of the $(3^3 \cdot 4^2)$ lattice 
of type $(33344c)$ and width $L_y=4$. 
The notation is the same as in Table \ref{table34a}.}}

\begin{center}
\footnotesize
\begin{tabular}{|cccc|} \hline
$J,I$ & planarity & does $\lambda_5$ enter? & features of ${\cal B}$\\
\hline\hline
${\bf v},{\bf v}$     & P,P & N & arcs \\
${\bf v_i},{\bf v_i}$, $i=1,2,3,4$ & NP,NP   & Y & arcs and one enclosed 
region \\
${\bf v_i},{\bf v_i}$, $i=5,6,7$   & P,P & N & arcs \\
${\bf v},{\bf v_i}$, $i=1,2,3,4$ & P,NP  & N & arcs \\
${\bf v},{\bf v_i}$, $i=5,6,7$ & P,P  & N & arcs \\
${\bf v_1},{\bf v_i}$, $i=2,3,4$ & NP,NP  & Y & arcs and one enclosed
region \\
${\bf v_1},{\bf v_i}$, $i=5,6,7$ & NP,P  & N & arcs \\
${\bf v_2},{\bf v_i}$, $i=3,4$ & NP,NP  & Y & arcs and one enclosed region \\
${\bf v_2},{\bf v_i}$, $i=5,6,7$ & NP,P  & N & arcs \\
${\bf v_3},{\bf v_4}$ & NP,NP  & Y & arcs and one enclosed region \\
${\bf v_3},{\bf v_i}$, $i=5,6,7$ & NP,P  & N & arcs \\
${\bf v_4},{\bf v_i}$, $i=5,6,7$ & NP,P  & N & arcs \\
${\bf v_5},{\bf v_i}$, $i=6,7$ & P,P  & N & arcs \\
${\bf v_6},{\bf v_7}$ & P,P  & N & arcs \\
\hline
\end{tabular}
\normalsize
\end{center}
\label{table34c}
\end{table}

\pagebreak

\begin{figure}
\vspace{-4cm}
\centering
\leavevmode
\epsfxsize=3.0in
\begin{center}
\leavevmode
\epsffile{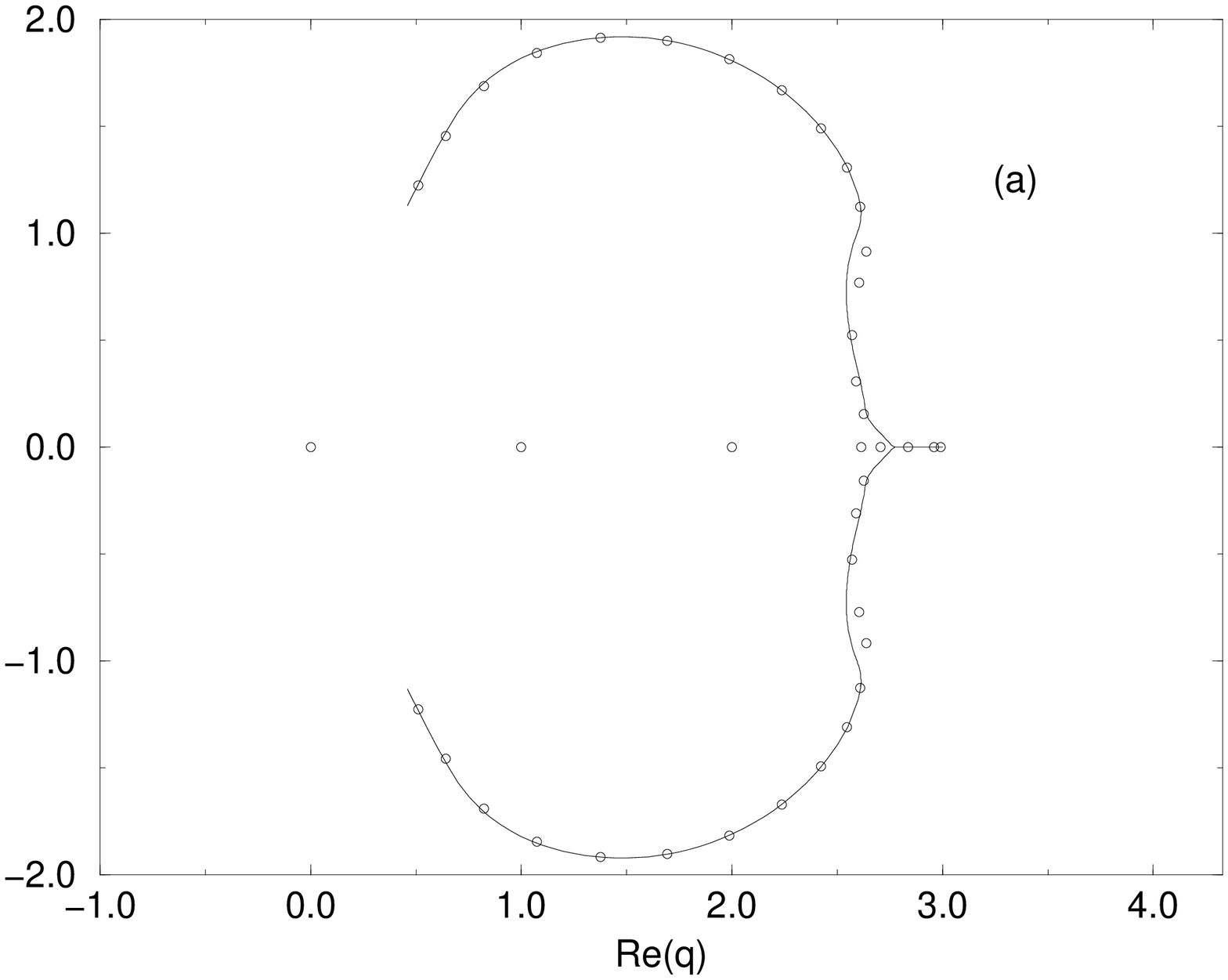}
\end{center}
\vspace{-3cm}
\begin{center}
\leavevmode
\epsfxsize=3.0in
\epsffile{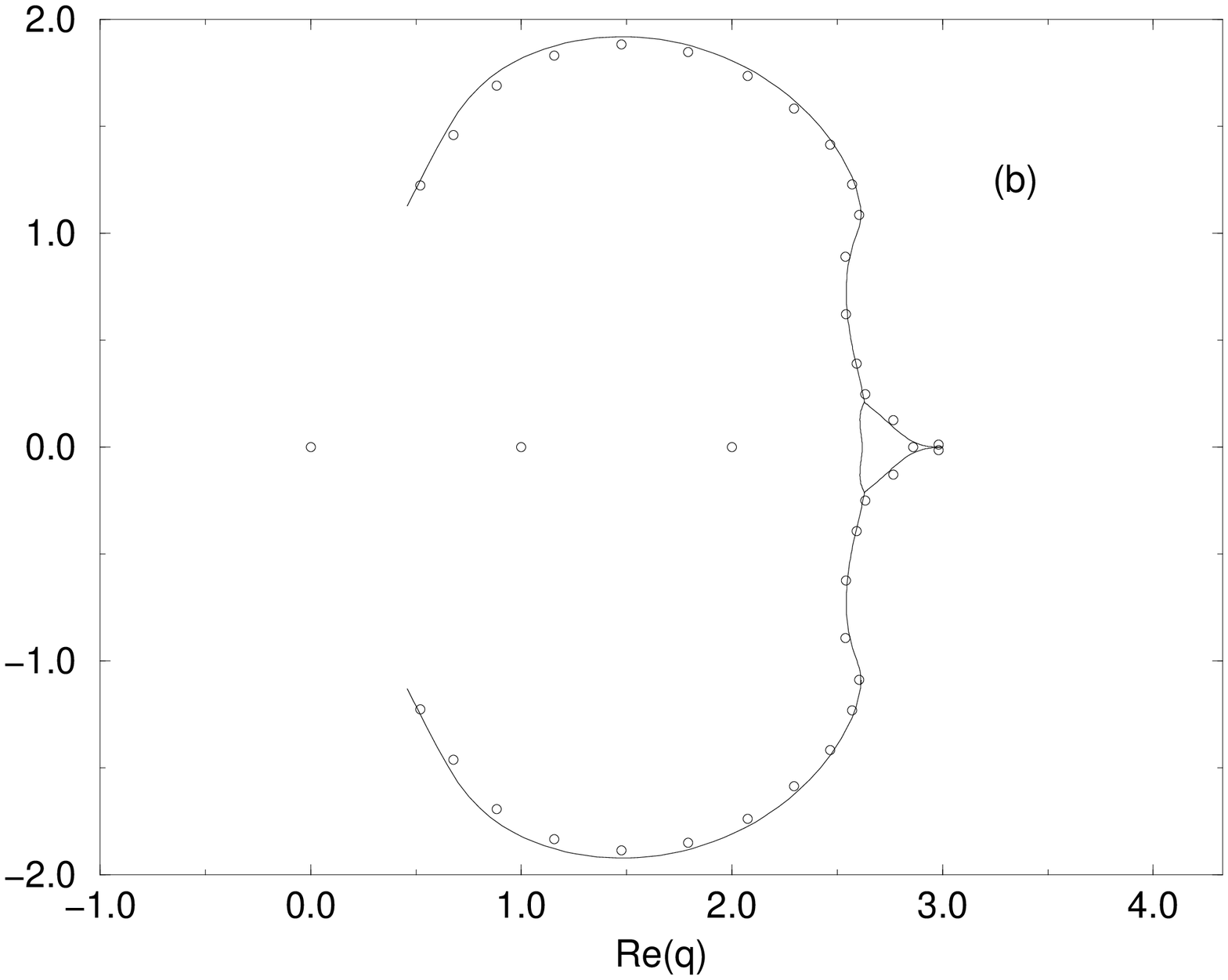}
\end{center}
\vspace{-2cm}
\caption{\footnotesize{Analytic structure of the function 
$W(\{G_{33344c(4)}'\},q)$ for a strip of the $(3^3 \cdot 4^2)$ lattice of type 
$(33344c)$, width $L_y=4$ and length $L_x=\infty$.
End subgraphs in (a) [(b)] are such that $\lambda_5$ does not [does] 
appear in the generating function.  For comparison, zeros of the chromatic 
polynomial for a strip with $m=8$ repeating units ($n=40$ vertices in (a) and
$n=36$ vertices in (b)) and end subgraphs 
described by (a) ${\bf v_6}$ [(b) ${\bf v_2}$] on the 
right and left ends of the strip, are shown. }}
\label{strip34c}
\end{figure}

\begin{figure}
\centering
\leavevmode
\epsfxsize=3.5in
\begin{center}
\leavevmode
\epsfxsize=3.5in
\epsffile{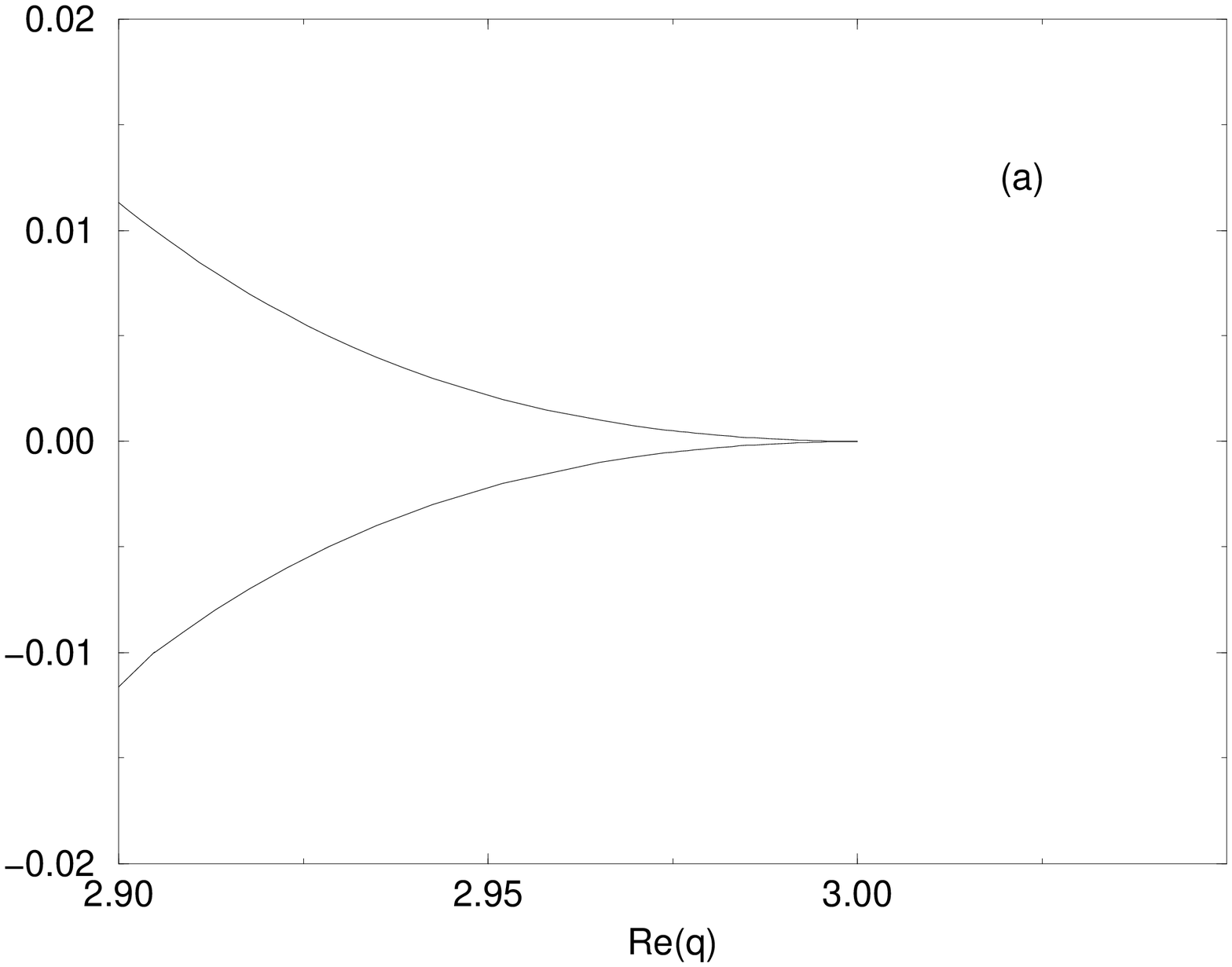}
\end{center}
\vspace{-4cm}
\begin{center}
\leavevmode
\epsfxsize=3.5in
\epsffile{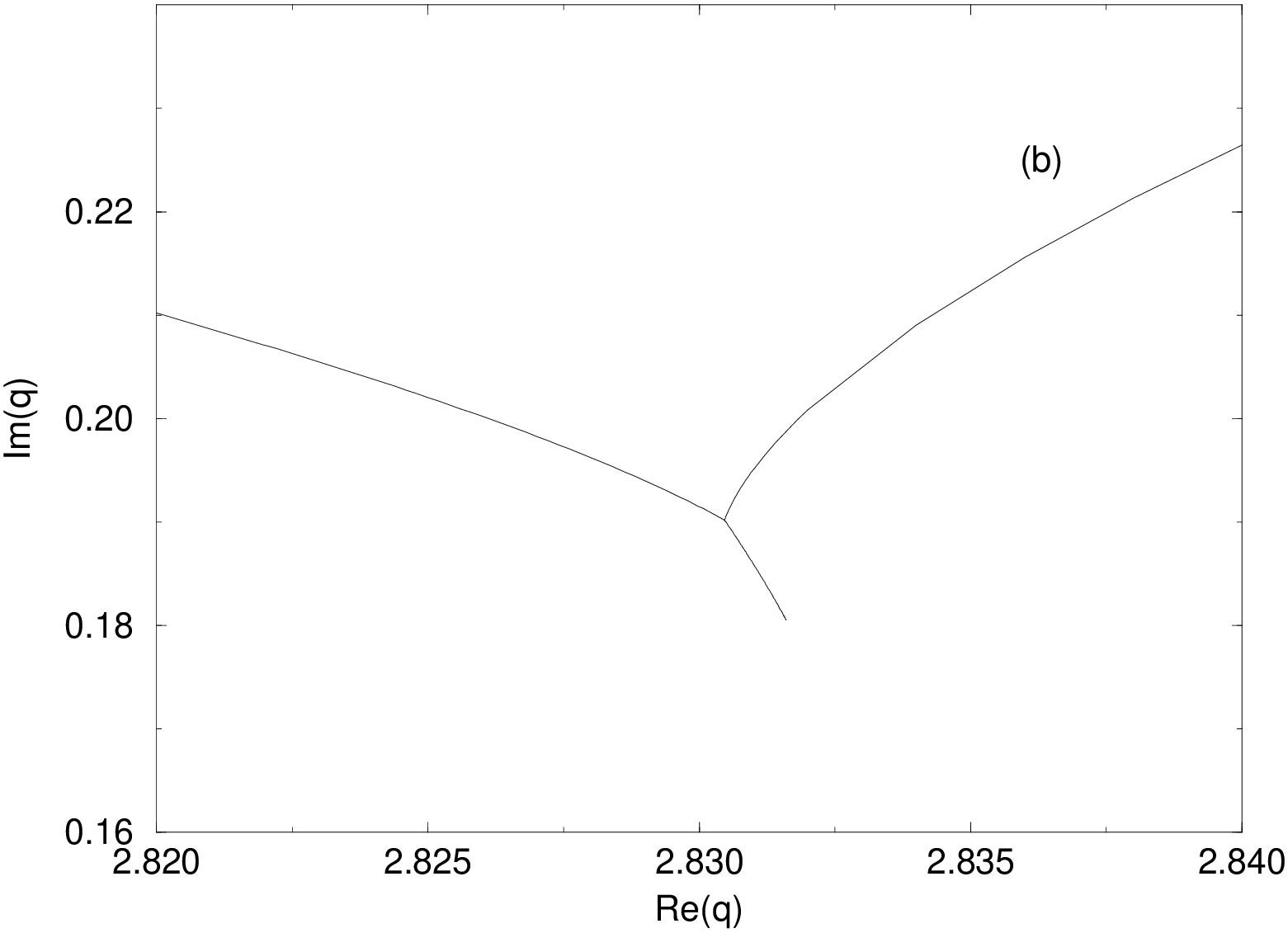}
\end{center}
\vspace{-2cm}
\caption{\footnotesize{Enlargement of the region in the complex $q$ plane
near the (a) cusp in Fig. \ref{strip34c}(b) and (b) multiple point with 
positive imaginary part in Fig. \ref{fig34btb}(a).}}
\label{fig34enlarge}
\end{figure}

Figs. \ref{strip34c}(a) and \ref{strip34c}(b) show the analytic structure 
of the $W(\{G_{33344c(4)}'\},q)$ function for a strip of the $(3^3 \cdot 4^2)$ 
lattice of type 
$(33344c)$ and width $L_y=4$. Fig. \ref{strip34c}(a) corresponds to end
subgraphs that 
do not yield $\lambda_5$ in the generating function. In these cases 
${\cal B}$ is formed by a self-conjugate arc and a line segment on the real
$q$ axis, extending from the multiple point $q\simeq 2.772$ to $q=3$. Fig. 
\ref{strip34c}(b) 
corresponds to end subgraphs that include $\lambda_5$ in the generating
function. In these cases ${\cal B}$ is comprised of arcs which enclose one 
self-conjugate region, where $\lambda_5$ is leading. The locus ${\cal B}$
crosses the real $q$ axis at $q=\frac{3+\sqrt{5}}{2}=2.618...$ and at $q=3$.
At the latter point the boundary ${\cal B}$ forms a cusp, as shown in Fig.
\ref{fig34enlarge}(a). At the point $q=3$ the denominator of the generating
function for the cases which yield boundary ${\cal B}$ as shown in Fig.
\ref{strip34c}(a), has the factored form ${\cal D}=(x-1)^4$. Similarly, for the
cases which yield boundary ${\cal B}$ as shown in Fig. \ref{strip34c}(b) 
the denominator factors as ${\cal D}=(x-1)^5$ at $q=3$.  For this type of strip
graphs all subgraphs on the two ends of the strips yield $q_c=\chi=3$. Another 
example of a boundary ${\cal B}$ which includes cusps was discussed in 
Ref. \cite{baxter87}.

\vspace{3 mm}

\section{Strip of the $(3^3 \cdot 4^2)$ lattice with periodic boundary
conditions in the transverse direction }

For a strip of the $(3^3 \cdot 4^2)$ lattice of type $(33344b)$ with periodic
boundary conditions in the transverse direction, transverse cross sections
forming squares, and subgraphs
on the two free ends of the strip described by the vector
${\bf v}$, we obtain a generating function with $j_{max}=2$, $k_{max}=3,$ and
coefficients
\beq
a_{33344(4),c,0}=q(q-1)(q-2)(q^5-11q^4+52q^3-134q^2+189q-115),
\label{a034tb}
\eeq
\beq
a_{33344(4),c,1}=-q(q-1)(q-2)(2q^7-30q^6+195q^5-716q^4+1613q^3-2239q^2+1775q
-617),
\label{a134tb}
\eeq
\beq
a_{33344(4),c,2}=q(q-1)(q-3)(q^2-3q+3)(q^3-7q^2+16q-11)(q-2)^4,
\label{a234tb}
\eeq
\beq
b_{33344(4),c,1}= -q^4+10q^3-43q^2+93q-82,
\label{b134tb}
\eeq
\beq
b_{33344(4),c,2}=(q-3)(2q^5-22q^4+100q^3-235q^2+284q-139),
\label{b234tb}
\eeq
\beq
b_{33344(4),c,3}=-(q-3)(q^3-7q^2+16q-11)(q-2)^4.
\label{b334tb}
\eeq

Three of the four eigenvalues of the $MD$ matrix are the inverses of the roots
of $1+b_{33344(4),c,1} x+b_{33344(4),c,2} x^2 +b_{33344(4),c,3} x^3$. These
three eigenvalues, denoted $\lambda_1$, $\lambda_2$ and $\lambda_3$,
contribute to the generating functions  for all subgraphs that
we consider on the two ends of the strip. The other eigenvalue is
$\lambda_4=1$, which contributes to the generating functions for some
subgraphs. In Table \ref{table34btb} we show for some
illustrative subgraphs on the two ends of the strips, whether or not
$\lambda_4$ contributes to the generating function and some
features of the boundaries ${\cal B}$ for each case.

\pagebreak

\begin{table}
\caption{\footnotesize{
Illustrative subgraphs for a strip of the $(3^3 \cdot 4^2)$ lattice
of type $(33344b)$ with periodic boundary conditions in the transverse
direction and transverse cross sections forming squares. The second column
indicates whether or not $\lambda_4$ contributes to the generating function.
The third column shows some features of ${\cal B}$. Notation is the same as in
Table \ref{table34a}}}

\begin{center}
\footnotesize
\begin{tabular}{|ccc|} \hline
$J,I$ & does $\lambda_4$ enter? & features of ${\cal B}$\\
\hline\hline
${\bf v},{\bf v}$   & N & arcs \\
${\bf v},{\bf v_i}$, $i=1,..,6$   & N & arcs \\
${\bf v_1},{\bf v_i}$, $i=1,2,3,4$ & Y & arcs and one enclosed region \\
${\bf v_1},{\bf v_i}$, $i=5,6$  & N & arcs \\
${\bf v_2},{\bf v_i}$, $i=2,3,4$ & Y & arcs and one enclosed region \\
${\bf v_2},{\bf v_i}$, $i=5,6$  & N & arcs \\
${\bf v_3},{\bf v_i}$, $i=3,4$ & Y & arcs and one enclosed region \\
${\bf v_3},{\bf v_i}$, $i=5,6$  & N & arcs \\
${\bf v_4},{\bf v_4}$ & Y & arcs and one enclosed region \\
${\bf v_4},{\bf v_i}$, $i=5,6$  & N & arcs \\
${\bf v_5},{\bf v_i}$, $i=5,6$  & N & arcs \\
${\bf v_6},{\bf v_6}$  & N & arcs \\
\hline
\end{tabular}
\normalsize
\end{center}
\label{table34btb}
\end{table}

\begin{figure}
\centering
\leavevmode
\epsfxsize=3.5in
\begin{center}
\leavevmode
\epsfxsize=3.5in
\epsffile{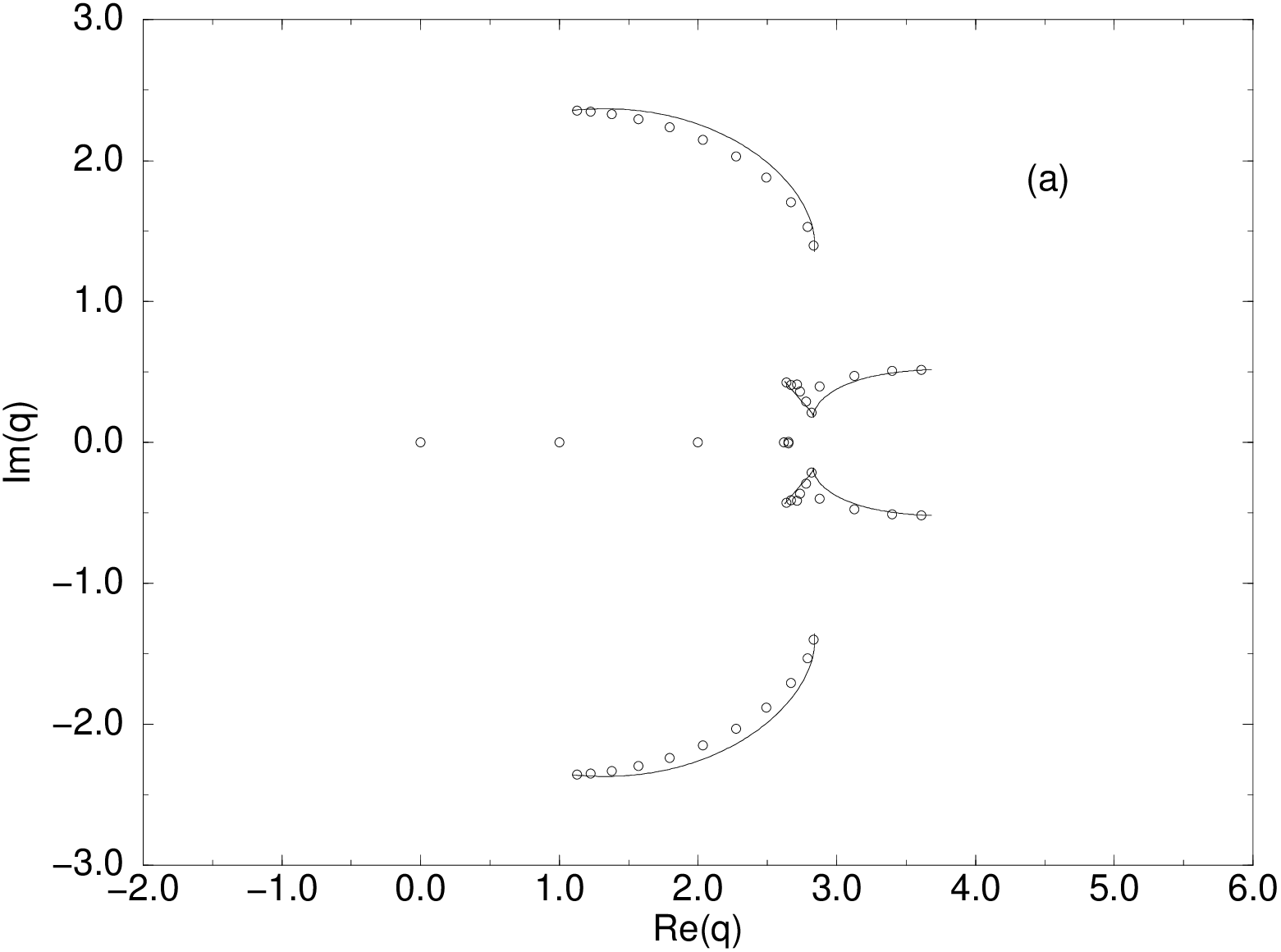}
\end{center}
\vspace{-4cm}
\begin{center}
\leavevmode
\epsfxsize=3.5in
\epsffile{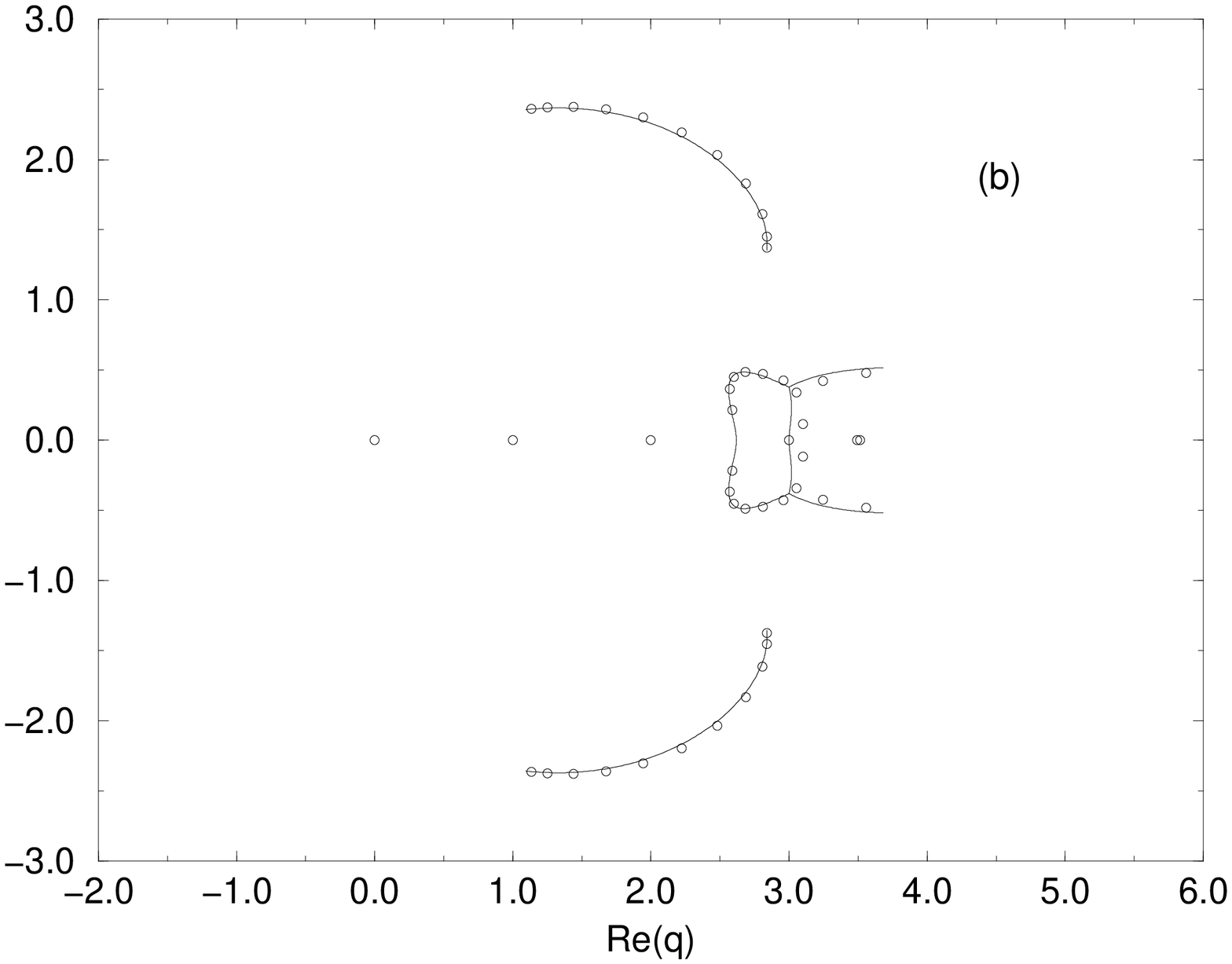}
\end{center}
\vspace{-2cm}
\caption{\footnotesize{Analytic structure of the function $W$ for the strip
of the $(3^3 \cdot 4^2)$ lattice of type $(33344b)$ with periodic boundary
conditions on the transverse direction, transverse cross sections forming
squares, and length
$L_x=\infty$. (a) corresponds to subgraphs on the free ends of the
strip for which $\lambda_4$ does not enter in the generating function. (b)
corresponds to cases where $\lambda_4$ enters in the generating function.
For comparison, the zeros of the chromatic polynomial for a strip with $m=10$
repeating units ($n=48$ vertices) and subgraphs described by (a)
${\bf v}$ [(b) ${\bf v_1}$] on the right and left ends of the strip, are
shown. }}
\label{fig34btb}
\end{figure}

For subgraphs where $\lambda_4$ is not present in the generating
function the analytic structure of the $W$ function, shown in
Fig. \ref{fig34btb}(a), is formed by two pairs of complex-conjugate arcs
which do not cross the real $q$ axis and do not enclose regions.
There is a pair of multiple points at $q\simeq 2.83\pm 0.19i$, which can be
seen more clearly in Fig. \ref{fig34enlarge}(b) where the region in the complex
$q$ plane around one of the multiple points is enlarged. 
For subgraphs where $\lambda_4$ is present in the generating
function the analytic structure of the $W$ function, shown in
Fig. \ref{fig34btb}(b), has arcs and a closed self-conjugate region, where
$\lambda_4$ is leading. In the cases where $\lambda_4$ is not present in the
generating function, the points $q \simeq 2.6243\pm 0.4328i$ correspond to
arc endpoints, whereas in the cases for which $\lambda_4$ contributes to the
generating function, these points lie within the enclosed region where
$\lambda_4$ is leading and do not correspond to arc endpoints.
The boundary ${\cal B}$ crosses the real $q$ axis at
$q=\frac{3+\sqrt{5}}{2}=2.618...$ and at $q_c=3$. 
It is interesting to note
that for a strip of the $(3^3 \cdot 4^2)$ lattice of type $(33344b)$ with
periodic boundary conditions on the transverse direction and transverse
cross sections
forming squares, in all cases where ${\cal B}$ crosses the real $q$ axis, we
have $q_c=3$, which coincides with $q_c$ for strips $G_{33344b(4)}'$ with
free boundary conditions on the longitudinal and transverse directions.
For both periodic and free boundary conditions on the transverse direction,
the boundary ${\cal B}$ only cross the real $q$ axis for certain subgraphs
on the longitudinal ends of the strip.

\section{Discussion and Conclusions }

In Ref. \cite{strip2} we have presented exact calculations of chromatic
polynomials $P$ and asymptotic limiting $W$ functions for strip graphs of 
the form $J(\prod_{\ell=1}^\infty H)I$, where $J$ and $I$ are 
various subgraphs on the left and right ends of the strip, and 
$(\prod_{\ell=1}^m H)$ are strips of various regular lattices consisting 
of $m$-fold repetitions of subgraph units $H$.  We have also studied the 
effects of using periodic as well as free boundary conditions in the 
transverse direction.  For the respective strip graphs we have determined 
the loci ${\cal B}$ where $W$ is nonanalytic in the complex $q$ plane.  
We find that for some $J(\prod_{\ell=1}^\infty H)I$ strip graphs, 
${\cal B}$ (i) does depend on $I$ and $J$, and (ii) can enclose regions in 
the $q$ plane. In the cases with free boundary conditions in the longitudinal
and transverse directions that were studied in Ref. \cite{strip2} the
boundary ${\cal B}$ enclosed regions when the end subgraphs $J$ and $I$
were non-planar graphs. In the present paper we present an example for this
class of strip graphs in which ${\cal B}$ encloses regions even for planar 
end graphs. The bulk of the specific strip graph that exhibits this 
property is a part of the $(3^3 \cdot 4^2)$ Archimedean lattice. The
boundary ${\cal B}$ for these strip graphs have multiple points and cusps.
We have also studied the effects of taking different types of strips of the
same uniform lattice as well as of using periodic and free boundary 
conditions in the transverse direction. It is interesting to note that
in all cases where ${\cal B}$ crosses the real $q$ axis, the maximum
crossing points are the same, namely $q_c=\chi=3$.  

\vspace{10mm} 

\begin{center}

Acknowledgments

\end{center}

Discussions with Professors Martin Ro\v{c}ek and Robert Shrock are gratefully
acknowledged. I would also like to thank Professor Robert Shrock for helpful
comments on the manuscript. This research was supported in part by the NSF 
grant PHY-97-22101.

\pagebreak

\vfill
\eject
\end{document}